\documentclass[doublecol]{epl2}

\title{Phase behavior of symmetric linear multiblock copolymers}
\shorttitle{Phase behavior of symmetric linear multiblock copolymers} 

\author{P. E. Theodorakis\inst{1,2,3} \and N. G. Fytas\inst{4}}
\shortauthor{P. E. Theodorakis and N. G. Fytas}

\institute{
  \inst{1} Faculty of Physics, University of Vienna, Boltzmanngasse 5, A-1090 Vienna, Austria, EU  \\
  \inst{2} Institute for Theoretical Physics and Center for Computational Materials Science, Vienna University of Technology, Hauptstra{\ss}e 8-10, A-1040 Vienna, Austria, EU \\
  \inst{3} Vienna Computational Materials Laboratory, Sensengasse 8/12, A-1090 Vienna, Austria, EU \\
  \inst{4} Department of Materials Science, University of Patras, 26504 Patras, Greece, EU
}
\pacs{36.20.-r}{Macromolecules and polymers}
\pacs{64.75.Yz}{Self-assembly}
\pacs{64.75.Gh}{Phase separation and segregation in model systems (hard spheres, Lennard-Jones, etc.)}

\abstract{Molecular dynamics simulations are used to study the
phase behavior of a single linear multiblock copolymer with blocks
of A- and B-type monomers under poor solvent conditions, varying
the block length $N$, number of blocks $n$, and the solvent
quality (by variation of the temperature $T$). The fraction $f$ of
A-type monomers is kept constant and equal to $0.5$, and always
the lengths of A and B blocks were equal ($N_{A}=N_{B}=N$), as
well as the number of blocks ($n_{A}=n_{B}=n$). We identify the
three following regimes where: (i) full microphase separation
between blocks of different type occurs (all blocks of A-type
monomers form a single cluster, while all blocks of B-type
monomers form another), (ii) full microphase separation is
observed with a certain probability, and (iii) full microphase
separation can not take place. For very high number of blocks $n$
and very high $N$ (not accessible to our simulations) further
investigation is needed.
 }

\begin{document}

\maketitle

The phase behavior of block copolymers has recently attracted much
interest in experimental and theoretical
studies~\cite{1,2,3,4,5,6,7,8,9,10,11,12,13,14,15,16,17,18,19,20,21,22,23,24,25,26,27,28,29,30,31,32,33,34,35,36,37,38,39,40,41,42,43,44,45,46}.
For melts of multiblock copolymer chains the phase diagram almost
resembles the phase diagram of that with diblock copolymer chains;
The  lamellar structure is expected for the most symmetrical case,
an approach which is particularly valid in the strong segregation
limit~\cite{33}. In the case of infinitely dilute solutions, it is
sufficient to consider the behavior of isolated chains, where
interactions (often of short range) within the chain are relevant.
The case of a single linear multiblock copolymer
chain~\cite{47,48,49,50,51} is very interesting even when only two
different types (A, B) of blocks of the same length are considered
(see Ref.~\cite{51} for discussion). Interestingly, multiblock
copolymer chains have also close relation to the various
toy-models (i.e., the HP model~\cite{52}) which try to mimic the
behavior of various biomacromolecules, which are formed by
periodically repeated structural units (``monomers'') along their
chain, in order to understand complicated biological phenomena,
i.e., protein folding~\cite{53}, helical structures~\cite{54},
etc.

Multiblock copolymers composed of two different types of blocks (A
and B) under good solvent conditions form coil structures and the
chain conformations are essentially dictated by the repulsive
interactions between the different monomers~\cite{47,48}. In fact,
the chemical difference of monomers should be kept responsible for
an expansion in the chain dimensions with respect to the
equivalent homopolymer chains (same total number of monomers)
under the same thermodynamic conditions~\cite{48,49,50}. Also, it
is expected that the spherical symmetry of the macromolecule
should be distorted adopting a slightly ellipsoidal overall
formation~\cite{48,49,50,51}. Nevertheless, the most interesting
behavior is expected when the solvent quality is varied. Under
poor solvent conditions the chain collapses, and different
scenarios of microphase separation between the monomers A and B
belonging to different blocks are conceivable depending on the
block length and the number of blocks of the chain at a given
temperature~\cite{47}. Identification of these different states
could be based on the analysis of the formed clusters as monomers
of the same, or different, blocks join together. For rather short
chains (accessible to simulations), simulation techniques provide
the best way to the understanding of the phase behavior of a
single multiblock copolymer, while a theoretical treatment could
be an elaborate task. To the best of our knowledge, the phase
behavior of a single linear multiblock copolymer has not yet been
discussed, even for the most symmetrical case.

In the following, we fill this gap and describe large-scale
molecular dynamics (MD) simulations of an off-lattice bead-spring
model of a single linear multiblock copolymer under poor solvent
conditions to provide a picture for the phase behavior of such
macromolecules. We remind the reader that one expects that for
such a simulated system sharp phase boundaries between different
states do not exist; rather smooth crossovers are expected.

\begin{figure}
\onefigure[width=3.2in]{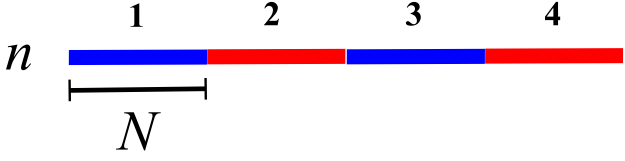} \caption{(Color online)
Definition of structural parameters describing our linear
multiblock copolymer chains. $n$ (in this case $n=4$) is the
number of different blocks A and B denoted with different color
(or grey tone) and $N$ is the length of each block. All the
blocks, irrespective of whether they are of type A or B, have the
same length $N$. Then the total length of the chains is $nN$.}
\label{fig1}
\end{figure}

In this study, we discuss results for the phase behavior of a
single symmetric multiblock copolymer with regular succession of
two different types of blocks (A and B) composing the linear
macromolecule. By using the term ``symmetric'' we mean that the
length of all blocks (A or B) was always equal ($N_{A}=N_{B}=N$)
and the total number of A ($n_{A}$) and B ($n_{B}$) blocks was $n$
($n=n_{A}+n_{B}$), with $n_{A}=n_{B}$ in all cases. Therefore, the
fraction of monomers of type A and B was always constant and equal
to $f=0.5$. A schematic picture of our parameters is given in
fig.~\ref{fig1}.

Then, our chains are modelled by the standard bead-spring
model~\cite{55,56,57,58}, where all beads interact with a
truncated and shifted Lennard-Jones (LJ) potential
\begin{equation}
\label{eq1}
U_{LJ}^{\alpha \beta}(r)= 4 \varepsilon_{LJ}^{\alpha \beta} [(\sigma_{LJ}^{\alpha \beta}/r)^{12} -
(\sigma_{LJ}^{\alpha \beta}/r)^6] + C, \quad r \leq r_c \quad,
\end{equation}
where $\alpha, \beta=A, B$ denote the different type of monomers,
and the constant $C$ is defined such that the potential is
continuous at the cut-off ($r_{c}=2.5$)~\cite{60}. For simplicity,
$\sigma_{LJ}^{\alpha \beta}=1$, $k_{B}=1$, but
$\varepsilon_{LJ}^{AA}=\varepsilon_{LJ}^{BB}=2\varepsilon_{LJ}^{AB}=1$, in
order to create an unmixing tendency between monomers A and B
belonging to different blocks as done in previous
studies~\cite{59,60,61}. Therefore, $\Delta \varepsilon=
\varepsilon_{LJ}^{AB}-1/2(\varepsilon_{LJ}^{AA}+\varepsilon_{LJ}^{BB})$ was
kept the same throughout our simulations. The connectivity along
the chain is maintained by the ``finite extensible non-linear
elastic'' (FENE) potential
\begin{equation}
\label{eq2}
U_{\rm FENE} =-\frac{1}{2} k r^2_0 \ln [1-(r/r_0)^2], \quad 0 < r
\leq r_0.
\end{equation}
$U_{FENE}(r\geq r_{0})= \infty$, and the standard choices
$r_0=1.5$, $k=30$ were used~\cite{55,56,57,58,59,60,61,62}.

\begin{figure}
\onefigure[width=3.2in]{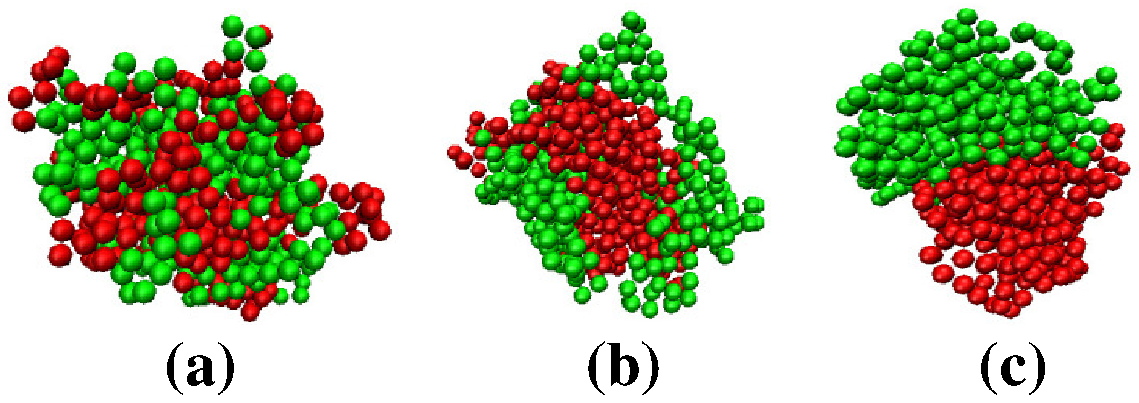} \caption{(Color online) Snapshot
pictures of three different multiblock copolymers of the same
total length $nN=600$ at a low temperature $T=1.5$ ((a): $N=6$,
(b): $N=15$, and (c): $N=60$). In case (c) we have the formation
for two clusters of different blocks which are always phase
separated. In case (a), full phase separation does not take place
and the number of clusters $N_{cl}$ has a symmetric variation
around an average value $2<N_{cl}<n$. The case (b) is the
intermediate case, where full microphase separation occurs with a
probability $P(N_{cl})>0$ ($N_{cl}=2\neq n$). Different colors (or
grey tone) correspond to different type of monomers (A,B).}
\label{fig2}
\end{figure}

For this model, the $\Theta$ temperature is known only rather
roughly, namely $\Theta=3.0$~\cite{62}. Being interested in $T
\leq \Theta$, we simulate temperatures ranging from $1.5$ to
$3.0$, where the chain collapses, and monomers of A and B blocks
cluster together with monomers of the same type and microphase
separation between A and B monomers takes place. Using standard MD
simulations~\cite{55,56,57,58,59,60,61}, the temperature was
controlled by the Langevin thermostat, i.e., the equation of
motion
\begin{equation}
\label{eq3} m_{LJ} \frac{d^2 \vec{r}_i}{d t^2 }= - \nabla
U_i-m_{LJ} \gamma \frac{d \vec{r}_i}{dt} + \vec{\Gamma}_i(t),
\end{equation}
for the time dependence of the monomer coordinates $\vec{r}_i(t)$
is numerically solved. Here $t$ is the time, $U_{i}$ is the total
potential the i-\emph{th} bead experiences and the friction
coefficient $\gamma$ was chosen $\gamma=0.5$ ($m_{LJ}$ is also
taken as unity for simplicity). The white noise
$\vec{\Gamma}_i(t)$ satisfies the standard fluctuation dissipation
relation, $\langle \vec{\Gamma}_i (t) \cdot \vec{\Gamma}_j (t')
\rangle = 6 k_BT \gamma \delta _{ij} \delta (t-t') $. To
equilibrate the systems close to the $\Theta$ temperature,
simulations of typically  $30 \times 10^6 \tau$ were carried out
at $T=3.0$, using a leap frog algorithm~\cite{63} with an
integration time step of $\Delta t=0.006 \tau$, where $\tau$ is
the natural time unit. After equilibration, we collect a number of
independent samples (typically $500$), which we use as initial
configurations for slow cooling runs. For longer chains,
temperatures higher than $T=3.0$ were used in order to facilitate
the procedure of obtaining initial independent samples. We note
here that the solvent is taken into account in our model only
implicitly by tuning the temperature as done in previous
work~\cite{55,56,57,58,59,60,61}. Then, decrease of the
temperature corresponds to higher incompatibility of the implicit
solvent with the monomers. For each cooling history, we lower the
temperature in steps of $\Delta T=0.1$ and simulate the system at
each temperature for a period that exceeds the relaxation time of
the chains, using the final configuration at each (higher)
temperature as starting configuration for the next (lower)
temperature. At low enough temperatures (typically $T<2.2$), where
dense ``clusters'' of a few neighboring blocks are formed, it is
not possible to run simple MD simulations long enough to sample
the phase space adequately. Therefore, using this procedure of
independent cooling histories is indispensable for obtaining
meaningful statistical results. This large statistical effort
prevents the study of exceedingly long chains, where also the
relaxation time of the chains becomes very high. We have also
performed runs where the temperature was changed in one step from
$T=3.0$ to the desired (lower) temperature, and after
equilibration, properties were calculated. While for $T\geq2.0$
the resulting properties did not show any significant dependence
on the different ``history of sample preparation'', somewhat more
pronounced correlations are seen for temperatures lower than
$T=2.0$ for a single run, which is an indication that the
intrinsic relaxation times begin to exceed the available
simulation time scale. Therefore, the use of independent cooling
histories was necessary in order to perform a meaningful
statistical analysis down to temperatures $T=1.5$.

\begin{figure}
\onefigure[width=2.5in,height=3.2in,angle=270]{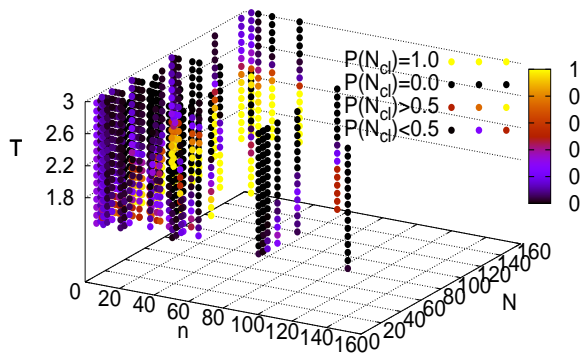}
\caption{(Color online) The probability of formation of a single
cluster of monomers A and a single cluster of monomers B in a
linear symmetric multiblock copolymer ($P(N_{cl})$, $N_{cl}=2\neq
n$) is plotted for different values of block length $N$, number of
blocks $n$ and temperature $T$. The different shading of colors
corresponds to different values of this probability.} \label{fig3}
\end{figure}

\begin{figure}
\onefigure[width=3.48in]{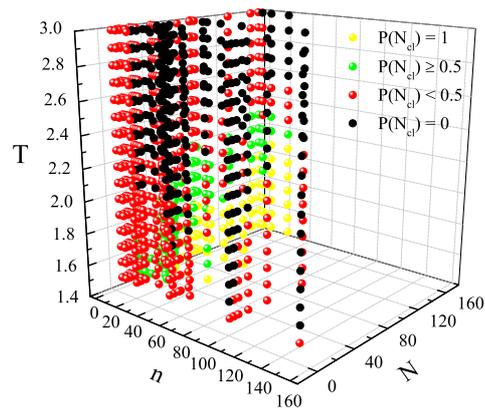} \caption{(Color online) Same as
in fig.~\ref{fig3}, but the colors are considered only according
to the four cases for the probability $P(N_{cl}),(N_{cl}=2\neq n$)
as indicated for clarity.} \label{fig4}
\end{figure}

When the chain collapses at the low temperatures and microphase
separated clusters of different blocks are formed, in order to
define the different types of phase behavior one needs to identify
for each configuration that is analyzed, which blocks belong to a
cluster. As in previous studies~\cite{64}, we have used the
standard Stillinger neighborhood criterion for monomers~\cite{65}.
When two monomers are less than a distance $r_{n}$ apart, they
belong to the same cluster, but monomers A and B are always
considered to belong to different clusters. We followed the
standard choice $r_{n}=1.5$, but we have also checked smaller
values for $r_{n}$ (i.e., $r_{n}=1.2$) and similar qualitative
results were obtained. Higher values of $r_{n}$ are physically
hardly significant, since then the monomers are too weakly bound,
due to the rapid fall-off of the LJ potential. fig.~\ref{fig2}
shows the different types of microphase separation that can be
observed in the case of symmetric multiblock copolymers. In case
(a), blocks of the same type can join together forming clusters,
but never all the blocks of the same type are able to form a
single cluster phase separated from another cluster of B monomers.
In this case, the number of formed clusters $N_{cl}$ fluctuates
symmetrically around an average value ($2<N_{cl}<n$). This
behavior is also an indication that the adopted simulation
protocol that we followed in this work leads to a meaningful
statistical analysis by considering this large number of
independent cooling histories, as is done in previous
work~\cite{56,59,61,64}. Moreover, analyzing separately the
properties for clusters of A monomers and for clusters of B
monomers we obtain the same results, which also shows the validity
of our simulation procedure. In case (c), all monomers A of all
different blocks belong to a single cluster of A-type monomers and
another cluster of B-type monomers is also formed. Then case (b)
is the intermediate case, where strong fluctuations in the number
of clusters  is observed and the occurrence of full phase
separation takes place with a probability $P(N_{cl})>0.0,
N_{cl}=2$. Thus, the different phase behavior in multiblock
copolymers can be characterized by defining the boundaries between
cases (a), (b), and (c) in terms of the probability $P(N_{cl})$
for $N_{cl}=2$.

figs.~\ref{fig3} and \ref{fig4} present results of the probability
of detecting full phase separation (i.e., all blocks of the A-type
monomers form a single cluster, while all monomers of B-type
belong to another cluster ($P(N_{cl})$, $N_{cl}=2$) between
monomers A and monomers B. The shading of the colors in
fig.~\ref{fig3} corresponds to this probability, whereas in
fig.~\ref{fig4} only the colors of the four different cases are
shown for clarity, without color shading according to the
probability $P(N_{cl}=2)$, as indicated in the figures.
Considering this probability for the full phase separation, we are
able to describe the different cases of the phase behavior, which
fig.~\ref{fig2} illustrates with characteristic snapshot pictures.
We have considered a broad variation of parameters, $n$, $N$, and
$T$, with $T \leq \Theta$, and we show here representative results
of our simulations for the description of the different regimes.
We can observe that for small block length $N$ ($N<20$), full
phase separation ($P(N_{cl}=2)=1$, as in case (a) of
fig.~\ref{fig2}) can be hardly seen independently of the number of
blocks $n$, even at low temperatures, where the chains collapse
and form globules. In this regime, the probability of full phase
separation is $P(N_{cl}=2)<0.5$. Looking more carefully in
fig.~\ref{fig3} we can actually observe that decrease of $n$
increases the probability of the formation of single clusters (one
with A monomers and another with B monomers), as expected.

The most interesting regime, where full microphase separation
occurs, is for rather small values of $n$, $4 \leq n \leq 20$ (of
course for $n=2$ the clusters of A and B monomers are considered
always separated). For this to happen, also the length of the
blocks $N$ has to be higher than a certain value ($N \geq 20$).
This regime is indicated in figs.~\ref{fig3} and fig.~\ref{fig4}
with yellow color (or the very light grey tone). It turns out that
the increase of $N$ for certain value of $n$ favors full phase
separation of A and B blocks, since for higher $N$ full separation
is observed, even at higher temperatures. In particular, for
$n=4,8,12$ and $N=150$ such an effect takes place even at
temperatures in the range $T=2.3$ to $T=2.4$. At high
temperatures, independently of $n$ and $N$, the probability
$P(N_{cl}) \approx 0$ ($N_{cl}=2$). In this case, the different
blocks prefer to be apart at high temperatures and can form
clusters with other blocks with monomers of the same type only
occasionally, due to thermal fluctuations. To conclude, full phase
separation is favored when the number of blocks is low, and the
block length higher than a certain value ($N \approx 20$), for the
range of temperatures that we used in our simulations. We note
that the different states occur gradually and the boundaries
between the different states are to be considered as smooth
crossovers (indicated also by the results of fig.~\ref{fig3}).
However, we have observed with closer inspection of our data that
these crossovers become sharper with the increase of the block
length $N$.

Very interesting is also the phase behavior observed for high
number of blocks $n$, while $N$ remains finite. In this case, $n$
is so high that the occurrence of full phase separation, as we
have defined it already in our discussion, is rather not observed
for the given set of potential parameters ($\Delta \varepsilon$).
What actually one sees is domains of different (A and B)
microphase separated clusters for the range of $N$ presented in
this study. It would be very interesting to simulate very long
block lengths, in order to check if full microphase separation is
possible for very high $N$ (or changing $\chi \sim \Delta
\varepsilon /T$; changing either $\chi$ or $N$ would result in
tremendous difficulties in simulating such a system) keeping also
high the number of blocks $n$. However, if one naively tries to
extrapolate our data for $n$ and $N$ in the range where full
microphase separation occurs, it could be suggested that full
phase separation for high $n$ could require also very high values
of $N$. It has also been discussed for copolymer melts~\cite{4},
that the geometry of the microphase separated regions is
controlled by the number of blocks $n$, as well as other
parameters, i.e., relative size and arrangement of the blocks.
Nevertheless, one would expect in the long chain limit in
equilibrium that a ground-state type structure would be a single
lamellar domain, where an interface between all A-type and B-type
blocks is formed, similarly to what is known for multiblock
copolymer melts~\cite{33}. Such a structure would have much less
(unfavorable) A-B contacts rather than a multidomain structure of
A and B clusters, which is kinetically favored due to the
preparation protocol in the simulation and the chosen set of
parameters accessible to simulations.

In summary, we have discussed the microphase separation of
symmetric linear multiblock copolymers, where a chain contains
blocks of monomers of types A and B, which alternate along the
chain via MD simulations of an off-lattice bead spring model. The
fraction of monomers A ($f_{A}$) and monomers B ($f_{B}$) was
always constant at $f=f_{A}=f_{B}=0.5$. Varying the number of
blocks $n$, the block length $N$, and the temperature $T$, we were
able to detect the regimes where full phase separation (all blocks
of type A belong to a single cluster and all blocks of type B to
another cluster), as well the regime that full phase separation
does not take place. We found that rather small values of $n$ ($4
\leq n \leq 20$) and increasing block length $N$ ($N > 20$) favors
full separation of the blocks, but $N$ should in no case be lower
than the aforementioned value. Let us point out here that, the
behavior of multiblock copolymers can be parallelized with that of
various biological macromolecules, which are formed by
periodically repeated structural units (``monomers'') along the
chain. In such systems the capability of the formation of clusters
could be related to the probability of forming bonds and
occurrence of interactions between different blocks on the same
chain. Further investigation is needed for very high number of
blocks and very high block lengths, not accessible to our
simulations.

\end{document}